# TUNEABLE CAPACITOR BASED ON DUAL PICKS PROFILE OF THE SACRIFICIAL LAYER

*Sofiane Soulimane, Fabrice Casset, François Chapuis, Pierre Louis Charvet, Marc Aïd*

CEA – Leti MINATEC, 17 Rue des Martyrs, 38054 GRENOBLE Cedex 9 (France)

## ABSTRACT

In this work, we describe a simple 1-mask sacrificial layer process that allows us to prototype a tuneable capacitor. The process is specially optimized to procure a dual picks profile of the sacrificial layer. The geometry design and the influence of the structural material stress were considered through the finite element analysis. The FEM simulation with Ansys was used to study in detail profiles of the sacrificial layer for the release of the membrane. Our approach is focused on Micro Electro Mechanical System capacitors. This is an emerging technology with a demonstrated potential for a wide tuning range tuneable capacitors and high quality factor.

## 1. INTRODUCTION

During the last decades, Microsystems have been a rapidly growing field with great future potential. The realisation of micro-electro-mechanical systems (MEMS) offers a high level of material integration. It requires simulation to allow the optimization of device performance before costly and time-consuming prototyping. We have seen real advancement in communications and RF technology. We consider tuneable capacitor as a key component in many radio frequency (RF) applications such as tuneable filters, voltage controlled oscillators (VCO), tuneable Low Noise Amplifiers (LNA) or Power Amplifiers (PA). It could be used for portable phones, computers or any communication systems. This study has been investigated to obtain high quality factor and wide tuning range tuneable capacitors. Typically, we research a demonstrator with capacitance value ranging from sub-Pico farad up to some Pico farad. In addition we look for a tuning range in excess of two hundred percent with a quality factor larger than fifty. This class of device has a high interest.

The capacitance value between two plates can be express by the following equation.

$$C = \frac{\varepsilon S}{d} \qquad \text{Equation 1}$$

Taking a look at equation 1, we can notice that we can obtain a capacitance variation with surface (S), dielectric (ε) or gap (d) variation.

Many references deal with the development of such components. Surface variation [1], dielectric variation [2] or gap variation components [3] are detailed in the state of the art. On these last devices, it is well known that the continuous tuning ratio of a conventional parallel-plate capacitor with electrostatic actuation is limited to 1.5:1, due to the so-called pull-in effect. Indeed, the displacement (x) of the beam induces an opposite spring (k) force on the beam described by equation 2.

$$F_{electrostatic} = \frac{\varepsilon S V^2}{2(d-x)^2} = kx \qquad \text{Equation 2}$$

When the displacement of the movable electrode reaches the third of the initial gap (displacement x≥d/3) the electrostatic force becomes greater than the spring force and the suspended beam will make contact with the bottom electrode ("pull-in effect").

Nevertheless, the electrostatic actuation seems to be the most promising way to obtain low consumption devices.

## 2. DUAL GAP TUNABLE CAPACITORS

In order to increase the tuning range of electrostatically actuated devices from an industrial perspective, dual gap designs are introduced [4, 5, 6].

The mechanisms of such component consist in the dissociation of the actuated and the capacitance areas by the use of two gaps. We design a component with an actuated gap ($E_a$) three times higher than the capacitive gap ($E_c$), as shown on Figure 1. By this way, to cover the linear tuning range we do the electrostatic actuation only on the third of the initial actuated area gap, without pull-in effect. This actuation induces a linear capacitance variation on the whole capacitance gap.

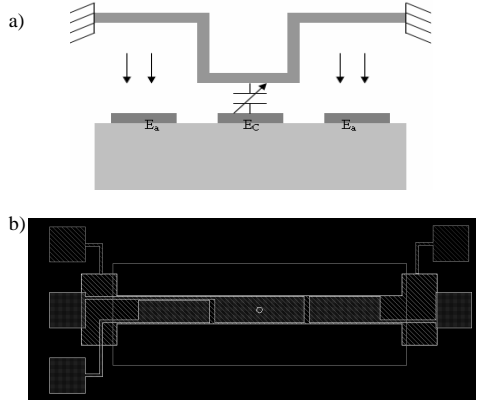

Figure 1: Schematic representation of dual gap tuneable capacitor - a) Cross section - b) Layout top view.

Process flow used in the state of the art to realize dual gap components induces multi-mask operations to pattern the fixed electrode with a dual thickness [5] or to pattern the sacrificial layer for the dual gap creation [4].
In this work, we developed a dual pick profile process by sacrificial layer treatments. We aim to obtain a device with the following specifications after release (Table 1).

|                  | Gap (µm) | Length (µm) | Width (µm) |
|------------------|----------|-------------|------------|
| Capacitive area  | 1.5      | 250         | 80         |
| Actuation area   | 4.5      | 200×2       | 80         |
| Complete beam    | Variable | 800         | 80         |

Table 1: Dual gap variable capacitor specifications.

With these specifications, we can predict the tuneable capacitor performances. Indeed, the mechanism of the dual gap tuneable capacitor induces a displacement of the capacitive area on the whole gap. We can estimate the maximum tuning range with the following hypothesis:

- Neglecting fringing capacitance or stress
- Including the use of dimples to stop the movable electrode before electrode to electrode contact (residual gap of 0.1µm).

By this way, we determine a tuning range higher than 1300% (Equation 3).

$$C = \frac{\varepsilon_0 S}{(d-x)}$$

$$C_{0\ (x=0)} = \frac{8.854.10^{-12} \times 250.10^{-6} \times 80.10^{-6}}{1.5.10^{-6}} = 0.12\,pF$$

$$C_{max\ (x=d-0.1)} = \frac{8.854.10^{-12} \times 250.10^{-6} \times 80.10^{-6}}{0.1.10^{-6}} = 1.77\,pF$$

$$TR = \frac{C_{max} - C_0}{C_0} = 1375\%$$

Equation 3

Moreover, we can determine a 12V actuation bias with the pull-in voltage equation (Equation 4).

$$V_{PI} = \sqrt{\frac{8kd^3}{27\varepsilon S_a}} = \sqrt{\frac{8\left(\frac{192EI}{L^3}\right)d_a^3}{27\varepsilon S_a}}$$

Equation 4

In the next section, we demonstrate how a dual pick sacrificial layer profile allowing the dual gap tuneable capacitor was achieved.

## 3. PROCESS FLOW AND SACRIFICIAL LAYER REALISATION

Overall, the use of a polymer substance is favourable to a low cost technology. Usually, the surface quality of the sacrificial layer determines the mechanical performance of the membrane material. It is thus important to improve the quality of sacrificial layer surfaces. Thus, our aim is to realise a process based on photo resist-like sacrificial layers with a high thermal stability (around to 350-400°C). This technology is compatible with integrated circuit (IC) technologies and could induce a high planarity, necessary to control the mechanical properties of the released structures. For this, we developed a bi-layer patterned photo resist process.
The photo-resist sample used in this study was commercial I, G, H-line photo resist (G-line [λ=436 nm], H-line [λ=405 nm], I-line [λ=365 nm]): Shipley S1800, Micro Resist technology map 1200 series…, which is a mixture of a base polymer of cresol–novolak resin and an additive with naphtoquinonediazide as the photosensitive material. The photo resist was coated on a substrate using a spin-coating method to be patterned below the movable material (membrane or beam). However, for sacrificial layer realization, we adopted a process that combines a standard lithographic patterning and thermal treatment in $N_2$ atmosphere. It increased glass temperature ($T_g$) and thermal stability of the resist layer ($T_g$). This is dependent on the viscoelastic materials properties, and so varies with the rate of applied load. We can consider this temperature as the limit temperature of the membrane deposition to have a stable profile of sacrificial layer and thermal stability for successor's steps.

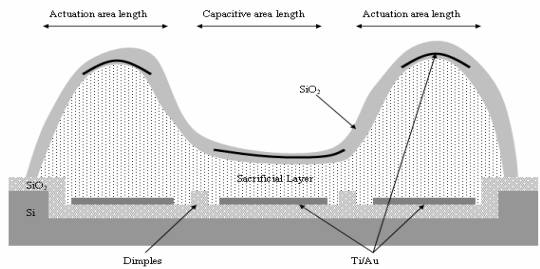

Figure 2 : Schematic representation of the tuneable capacitor based on dual pick profile of the sacrificial layer

The main idea of our approach is based on a special profile of the sacrificial layer designed by two picks (Figure 2). This process is inspired from an earlier study presented in [7]. Wherein, we study planarity of novolak photo resist to realise a structural membrane. The sacrificial process requires a thermal cycle whose generate two picks on the edges of the patterned layer at 115-120°C (figure 3b). Their sizes increases as function as temperature increase (figure 3c).

This growth ends at ~320-350 °C, and a smooth slope on the edges of the patterned structure can be observed. A reverse thermal cooling is required to save the final profile. We can denote three main conclusions about this phenomenon:

- In first, we obtain a two-peaks profile for large motifs (Figure 4a), and a smooth planar profile for small motifs presumably because the two peaks merge (Figure 4b)
- Secondly, we remark that the transition between these two profiles depend on both resist thickness. More resist thickness increase as soon the two picks merge.
- In third, we observe the same phenomena for a resist processed on a cavity, and the transition length-scale was increased for the same thickness. Also, we can observe that more the cavity is too wide more the difference of thickness between two picks and centre of the motif is large.

Allowing these conclusions, we establish à tri-layer resist process to set in order the thickness of dual picks. The process flow includes first the realization of large motif cavity (4.5 µm wide and 800 µm large). Then we pattern the fixed electrode. The sacrificial layer is realised as follows: First, a sufficient thickness of resist was patterned and cured as described in [7]. The following step consists in an $O_2$ plasma etching with a full exposition of the wafer in order to trim the resist on the top of the electrodes. This step is required to avoid the roughness caused by electrodes.

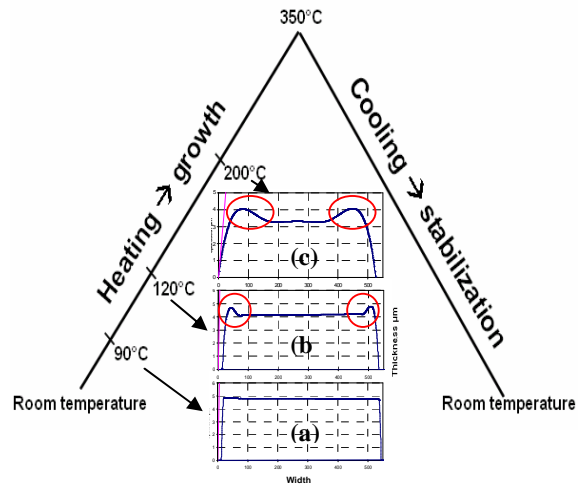

Figure 3: Thermal cycle presentation and Step 2D (Tencor profilometer) profiles the patterned resist as a function of the temperature: (a) 90°C, (b) 120°C and (c) 200°C..

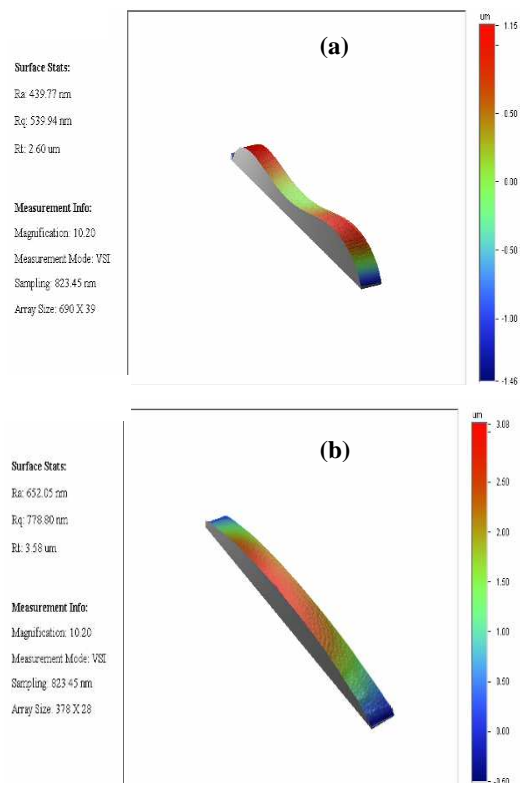

Figure 4 : 3D profile of the polymeric sacrificial layer with Veeco interferometer (a)Two picks profile for a large motif : 600 μm (b) planar and smooth for a small motif : 150 μm.

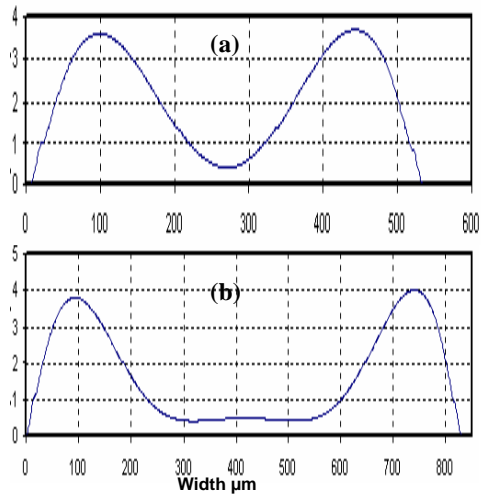

Figure 5: 2D profile with Tencor profilometer on 4.5 um cavity motif depth for (a) 520 μm and (b) 820 μm, motif large.

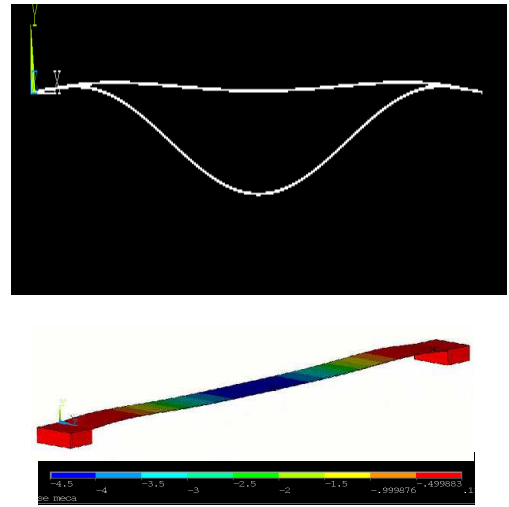

Figure 6: ANSYS FEM simulation – a) Release membrane profile (cross section) – b) 3D Release membrane profile.

Secondly, another resist layer (3 µm) is identically patterned and cured to procure the picks in the edge of the cavity. A third layer is required to growing the picks. Figure 5 shows final profile for 520 and 800µm long profile (4.5µm cavity depth). We note that larger is the motif, larger is the actuation area.

Then a succession of SiO2 / gold deposit and etch steps permit to realize the suspended beam and electrode

Finally, an isotropic etching is required to release the sacrificial layer. It is well known that the removal of the highly photo-resist cured films treated with a wet etching without leaving any residues and avoid capillarity or sticking problem, is quite difficult. To remedy to this problem, down stream plasma (MATRIX machine) is used for sacrificial layer release.

After release, the residual stress in the membrane will induce the deflexion of the mobile part of the component. Utilization of cavity is required to avoid sticking problem of the membrane regarding this deflexion. The following section deal with the simulation of the membrane deflexion after release.

## 4. FEM SIMULATION

A variety of specialised tools is available in the MEMS design flow. FEM tools (ANSYS, Coventor, Comsol…) are widely used on component level to optimize the MEMS design.

We used ANSYS to study the dual picks membrane behaviours after release.

Indeed, residual stress in the structural material will induce a displacement of the free structure after sacrificial layer etch. For this, structural membrane encapsulates electrodes with $SiO_2$ moving parts are simulated.

Structures and profiles resulting on well known dual picks shapes have been designed using FEM simulations.

We simulated the behaviours of a dual pick membrane realized on this sacrificial layer after release. The membrane is an 800*80µm² beam. We can notice a 4.5µm deflexion of the centre of the beam due to the residual stress in the structural membrane material as shown on figure 5. The figure 5-a shows the profile of the released membrane with an exaggeration factor of 5. The figure 5-b illustrates this profile with a 3D view.

We obtain a residual gap of nearly 1.5µm for the capacitive area by this process, whereas we have more than 7 µm gap for the actuation area.

Compared to the desired profile, we can notice a more important actuation gap than expected. It will induce an increase of the actuation bias. With Equation 4 we can predict a 23V actuation bias with such initial actuation gap.

## 5. CONCLUSION

In this paper, we present a novel dual gap tuneable capacitor process based on the profile of the sacrificial layer. This profile involves a tri-layer photo-resist process with only one mask level inducing a two picks resist sacrificial layer profile. Regarding this dual gap sacrificial layer profile, we propose a MEMS tuneable capacitor having one movable plate loaded with a $SiO_2$ layer. We estimate with a simple model that this tunable capacitor could present a wide linear tuning range in excess of 1000% under 23V.

Important work is still needed both on the process and the design to obtain a low actuation voltage by decreasing the actuation gap. This membrane profile is obtained by the sacrificial layer control due to important work on the different parameters influence. The sacrificial layer control can now allow different new profiles for various applications.